\documentclass[aip,graphicx,numerical]{revtex4-1}
\usepackage{graphicx}%
\usepackage[colorlinks=true,linkcolor=blue]{hyperref}%
\usepackage{soul}
\usepackage{xcolor}
\draft 

\begin{document}


\title{Reproducibility and off-stoichiometry issues in nickelate thin films grown by pulsed laser deposition} 



\author{Daniele Preziosi}
\email[]{daniele.preziosi@thalesgroup.com}

\author{Anke Sander}
\author{Agn\`es Barth\'el\'emy }
\author{Manuel Bibes}
\noaffiliation
\affiliation{Unit\'e Mixte de Physique, CNRS, Thales, Univ. Paris-Sud, Universit\'e Paris-Saclay, 91767, Palaiseau, France.}


\date{\today}

\begin{abstract}
Rare-earth nickelates are strongly correlated oxides displaying a metal-to-insulator transition at a temperature tunable by the rare-earth ionic radius. In PrNiO$_3$ and NdNiO$_3$, the transition is very sharp and shows an hysteretic behavior akin to a first-order transition. Both the temperature at which the transition occurs and the associated resistivity change are extremely sensitive to doping and therefore to off-stoichiometry issues that may arise during thin film growth. Here we report that strong deviations in the transport properties of NdNiO$_3$ films can arise in films grown consecutively under nominally identical conditions by pulsed laser deposition; some samples show a well-developed transition with a resistivity change of up to five orders of magnitude while others are metallic down to low temperatures. Through a detailed analysis of \textit{in-situ} X-ray photoelectron spectroscopy data, we relate this behavior to large levels of cationic off-stoichoimetry that also translate in changes in the Ni valence and bandwidth. Finally, we demonstrate that this lack of reproducibility can be remarkably alleviated by using single-phase NdNiO$_3$ targets.
\end{abstract}


\maketitle 

\section{Introduction}
When embarking in a new study of a given material in thin film form, a primary step concerns the optimization of the growth procedure, initially to obtain samples with physical properties mimicking those of the bulk. Then, as demonstrated by plentiful studies in the literature\cite{growthmode2016} , the films' properties can be tailored by diversifying for instance, the deposition conditions, the substrate-induced strain\cite{Schlom2007} and the overall film thickness.\\
\textit{Per contra}, reports on the effective reproducibility of the particular electronic and/or magnetic properties, for identically repeated growth experiments, are not frequent and sometimes reproducibility is only assumed to be fulfilled \cite{Baker_2016} . This issue is particularly critical for multi-element compounds such as transition metal perovskite oxides. Thin films of these materials are usually grown by processes such as sputtering, molecular-beam epitaxy or pulsed laser deposition (PLD) in which a fine tuning of film composition can be very delicate because growth parameters are often interdependent. This is especially the case of PLD in which thin film deposition proceeds from the ablation of a ceramic pellet, a non-equilibrium and very complex process \cite{PLDeltit} .\\
Recently, the sharp metal-to-insulator transition (MIT) displayed in perovskite rare-earth nickelate oxides ($\textit{R}$NiO$_3$, $R$ being the rare-earth) has stimulated the interest of the oxide electronics community. Indeed, thin films of $\textit{R}$NiO$_3$ which crystallize in the distorted GdFeO$_3$-perovskite-like structure, represent an interesting playground for fundamental physics studies \cite{Medarde_1997} and also have attractive perspectives for prototype devices \cite{Scherwitzl_2010} . In nickelates, the MIT can be tuned within a very large temperature range by changing the ionic radii of the rare-earth ions ($\textit{R}$$\neq$La) \cite{Lacorre1991} . The MIT takes place simultaneously to the onset of an anti-ferromagnetic order of the Ni-3d spins along the [111] direction of the pseudocubic unit-cell for $\textit{R}$=Nd,Pr \cite{PhysRevB.46.4414} . The literature contains a plethora of experimental studies in which different routes have been employed to tune the transport properties of $\textit{R}$NiO$_3$ ceramics and thin films. In particular, the MIT appears to be very sensitive to chemical doping \cite{Garc_a_Mu_oz_1995, Lian_2013} , electric-field effect \cite{Scherwitzl_2010,Yang2011} and epitaxial strain\cite{ZhangStrainNNO, LiuStrainNNO, Scherwitzl_2010, PhysRevB-Bruno} . 
However, in the specific case of NdNiO$_3$ (NNO) thin films, studies on the effect of different strain states provide (unexpected) dissimilar results. For instance, compressive strain has been found to decrease, increase or completely suppress the MIT as summarized by Middey $et\,al.$ in their review (see, Ref.~\cite{Middey_2016} and references therein). This suggests that another parameter is at play to determine the transport response of the films.\\Recently, Breckenfeld $et\,al.$ have shown that the cation stoichiometry plays a crucial role in the transport properties for different strain states\cite{Breckenfeld_2014} . On the other hand, Hauser $et\,al.$ reported that the off-stoichiometry of the NNO films is effectively not affecting the temperature at which the MIT occurs but that, only the strain state needs to be considered\cite{Hauser_2015} . These contradictory results suggest that there is a problem of reproducibility of the physical properties of NNO thin films that are generally grown via physical vapor deposition techniques onto similar templates. \\
In this paper we report on the growth of NNO thin films by pulsed laser deposition, using two types of ceramic targets, \textit{i.e.} mixed and pure phases. In the former case, despite the absence of spurious phases in X-ray diffraction (XRD) spectra, and a good morphology as inferred by atomic-force microscopy (AFM) images and reflection high-energy electron diffraction (RHEED), the resistivity curves displayed extremely dissimilar properties for nominally identical samples. Using \textit{in-situ} X-ray Photoelectron Spectroscopy (XPS) we found that the degradation of the transport properties was correlated with a systematic change in the Nd/Ni peak area ratio. This change in thin film stoichiometry is likely associated with a poor control of the thermodynamic stability of the NdNiO$_3$ pure phase in the laser-induced plume. This problem was solved by using a pure phase target, with which we could produce NNO thin films with a much better degree of reproducibility.

\section{Experimental section}
Bulk NNO is indexed in the orthorhombic crystal structure ($Pbnm$ space group) with lattice parameters: $a$\,=\,5.389 \AA, $b$\,=\,5.382 \AA, and $c$\,=\,7.610 \AA (the pseudocubic lattice parameter is $a_{pc}$\,=\,3.807 \AA). The LaAlO$_3$ (LAO) substrates exhibit a rhombohedral crystal structure (R$\overline{3}$c space group) with $a_{pc}$\,=\,3.790 \AA. Hence, NNO thin films onto LAO(001) single crsytals (CrysTec GmbH) experience a compressive static strain of -0.45\%. The substrates were etched (during $ca.$ 4\,min) in a buffered HF solution and then annealed in oxygen for two hours at 1050$^{\circ}$\,C to achieve atomically smooth surfaces with step-terrace formation. Thin films of NNO were grown via pulsed laser deposition (KrF excimer laser) at a temperature of 700$^{\circ}$\,C in a pure oxygen environment (pressure of 0.2\,mbar). The base pressure of the chamber prior deposition was of the order of  10$^{-9}$\,mbar and the cooling was performed with a rate of 10$^{\circ}$\,C/min in an oxygen pressure of 1\,bar. The repetition rate was kept constant to the value of 2\,Hz while the fluence values were varied in the 0.8-2\,J/cm$^{2}$ range. For all growth experiments, the target-to-substrate distance was kept at the value of 4.5 cm and the laser spot area was ca. 0.02\,cm$^{2}$. The focused laser spot was rectangular in shape with very sharp edges. \\
Ceramic targets with a mixture of Nd$_{2}$O$_{3}$ and NiO phases from Pi-Kem (density 6.55\,g/cm$^{3}$) and with a stoichiometric phase of NdNiO$_{3}$ from Toshima (density 3.51\,g/cm$^{3}$) were used. Both targets were polished with a sandpaper (600-grit) to recover the surface after 6/7 growth experiments, hence, each NNO thin film was grown with the target having a different ablation history. Please note that the $\textit{R}$NiO$_3$ are a class of materials which are difficult to prepare in bulk\cite{Alonso2005} . Therefore, the commercially available ceramic targets contain a mixture of rare-earth- and Ni-oxides based compounds and, only for few materials single phase ceramic targets of relative high density exist.\\ $\textit{In-situ}$ XPS measurements using a Mg K$\alpha$ source ($h\nu$\,=\,1253.6 eV) were performed subsequently to the thin film growth. All spectra were measured in normal emission, $\textit{i.e.}$ photoelectron ejected perpendicular to the sample plane. Spectra analysis were carried out with the CasaXPS software. Transport measurements were performed with a Dynacool system from Quantum Design in the Van der Pauw configuration after sputtering top electrodes of gold. A current of 0.5\,$\mu$A was used to ensure the measure of the very high resistance state even at low temperatures. 

\section{Growth from a mixed-phases target}
\begin{figure}
  \includegraphics[angle=0,width=\columnwidth]{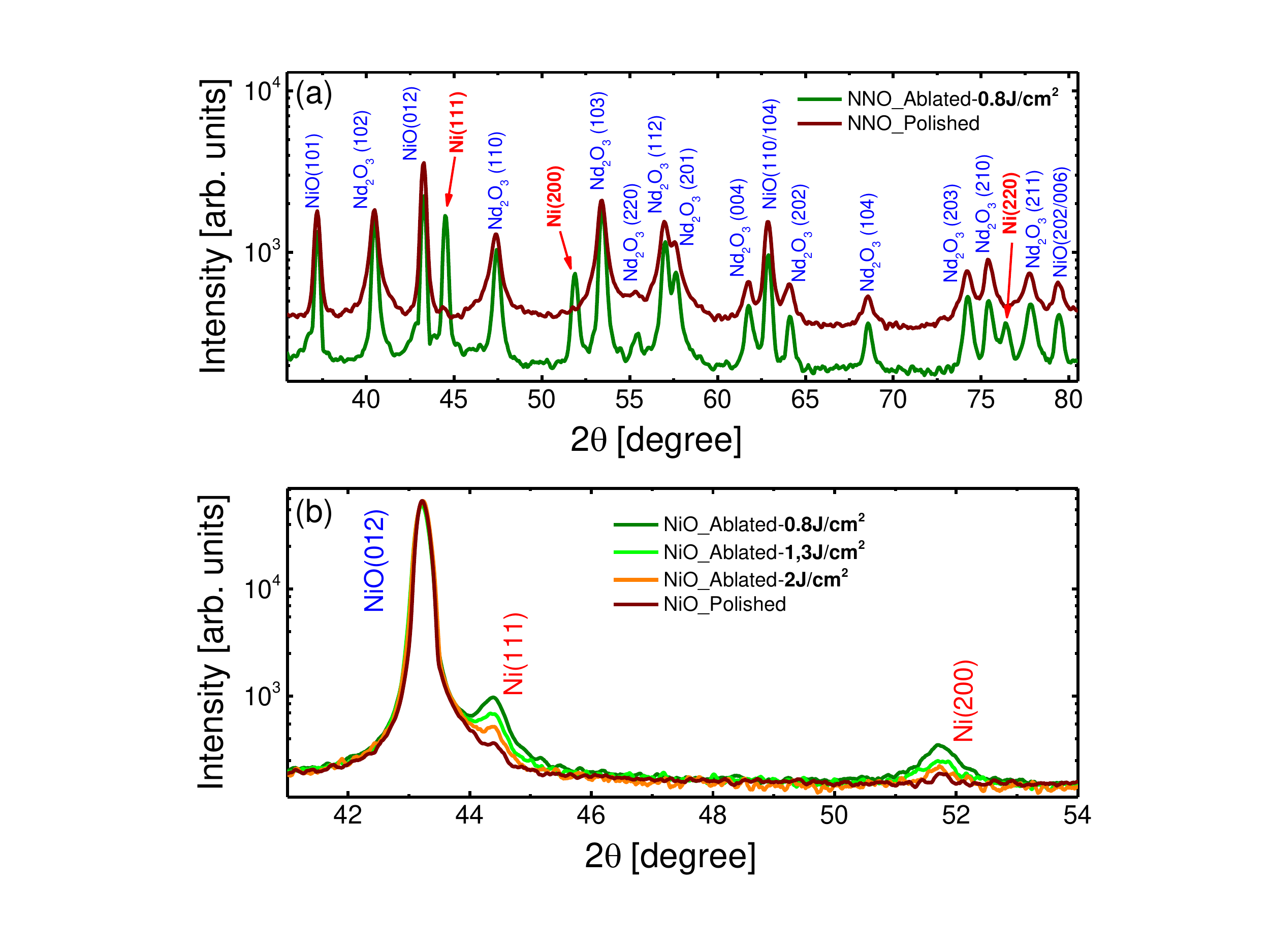}
  \caption{a) X-ray diffraction patterns of the mixed-phases target prior and after laser ablation. Metal Ni peaks appear after ablation ($cf.$ red arrows). b) X-ray diffraction patterns acquired onto a NiO oxide ceramic target as a function of the laser fluence. }
  \label{fig1}
\end{figure}
A valuable premise to reduce possible sources of off-stoichiometry \cite{Dam_1998} and, eventually, impose a good level of reproducibility of the oxide thin films, is the understanding of the influence of the ablation process on the chemistry of the target surface. Indeed, depending upon the element-specific ablation thresholds and the density of the ceramic target itself, differential ablation processes can give rise to unexpected segregation onto the target surface. As a result, the ablation history of the target ($i.e.$ the whole number of laser pulses fired onto it), can become a determinant factor for the reproducibility issue.\\  
In order to obtain this first important feedback for our growth experiments, we performed XRD measurements of the target after the polishing and the ablation procedures, respectively. Initially we fixed the fluence value at 0.8\,J/cm$^{2}$ and the obtained XRD patterns in a relatively large 2-theta range are shown in Figure \ref{fig1}a. Before and after ablation, the scans show reflections from both NiO and Nd$_2$O$_3$, as expected, but a segregation of Ni metal is detected after ablation ($cf.$ red arrows). This suggests that some of the NiO was reduced to metallic Ni through the interaction with the laser beam. We then addressed the influence of the laser fluence on the NiO reduction process. To preserve the NNO mixed phase target we used a simple NiO ceramic target. Figure \ref{fig1}b shows the evolution of the Ni metal diffraction peaks for various laser fluences, and the data indicate that the reduction of the NiO to Ni metal is promoted at low fluence values ($i.e.$ <\,2\,J/cm$^2$). 
\begin{figure}
  \includegraphics[angle=0,width=\columnwidth]{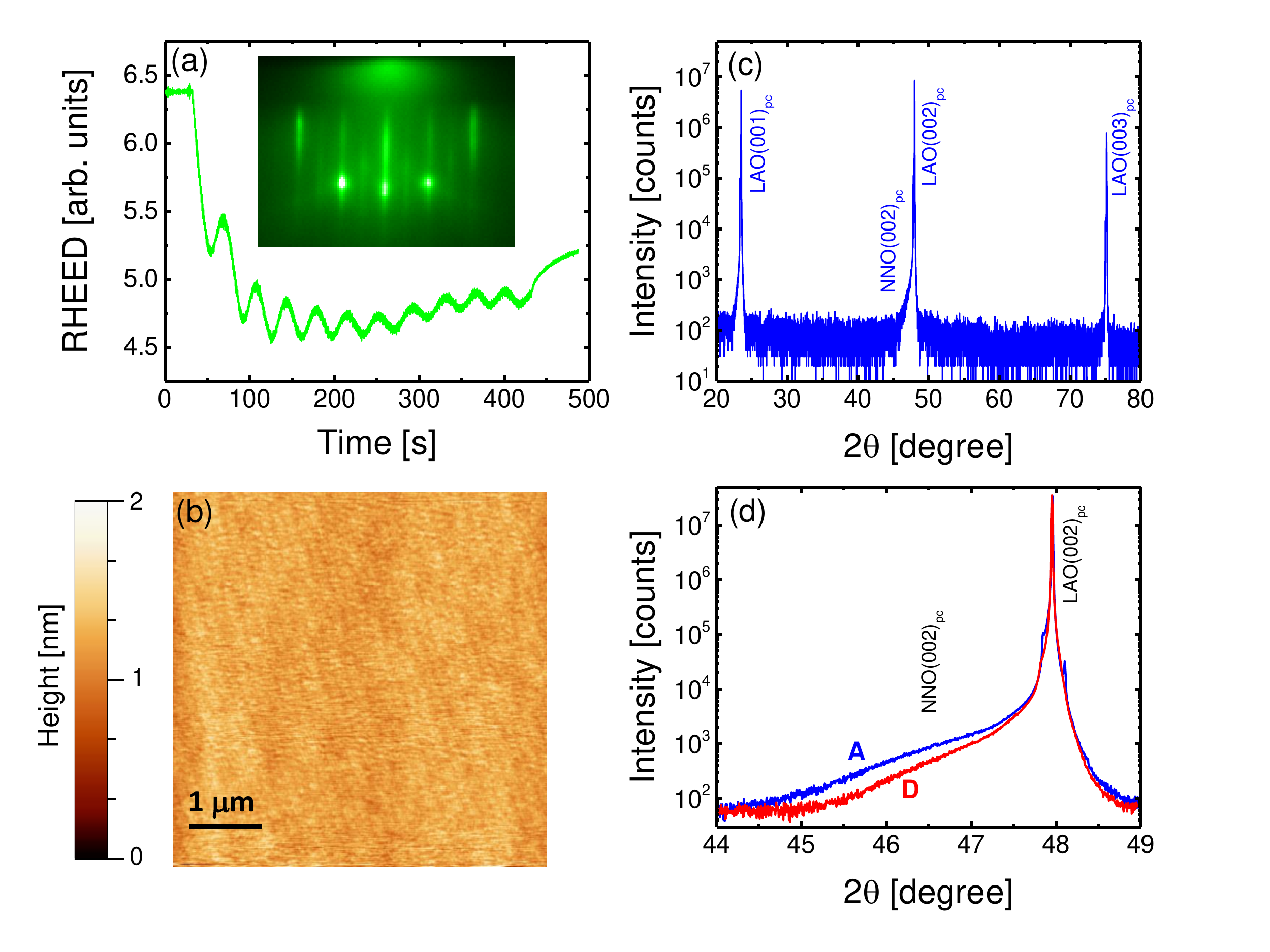}
  \caption{a) Representative RHEED oscillations characterizing the growth of NNO thin films with related streaky pattern in the inset. b) Atomic force microscopy image revealing a very smooth surface. c) Representative X-ray diffraction pattern where no spurious phases are present. d) Slight variation of the out-of-plane lattice parameter between two NNO thin films.}
  \label{fig2}
\end{figure}
As a result, a series of NNO thin film were grown at the relatively high fluence value of 2\,J/cm$^2$ using the mixed phase NNO target. For all films, RHEED patterns (Fig. \ref{fig2}a) showed well-defined spots, indicating a well-ordered film surface. We observed clear RHEED oscillations and set the film thickness to 11 unit cells for all samples ($cf.$ inset of Fig. \ref{fig2}a). A smooth sample surface was confirmed by atomic force microscopy (Fig. \ref{fig2}b). XRD revealed that all NNO thin films grew single phase and no spurious phases were observed (see, for example, Fig. \ref{fig2}c). \\
Using these conditions, we then proceeded to grow consecutively five nominally identical NNO thin films (labeled A, B, C, D, and E). From the RHEED and AFM analysis these films looked virtually identical. However, the position of their (00$l$) reflection varied between samples as illustrated in Figure \ref{fig2}d for two films. Because of the low thickness it proved impossible to perform reciprocal space maps, but both their layer-by-layer growth and their flat morphology suggest that they are all fully strained. The difference in the out-of-plane lattice parameters inferred from the positions of the (00$l$) peaks is thus likely to reflect a change in the unit-cell volume resulting from variations in compositions \cite{Breckenfeld_2014} . \\
In Figure \ref{fig3} we plot the resistivity curves as a function of the temperature obtained in a cooling-to-warming cycle for all samples. Surprisingly, very different behaviors are observed. All films show the MIT and the hysteretic behavior typical for NNO thin films \cite{Blasco_1994} except sample E that is metallic at all temperatures. Films grown after sample E before repolishing the target showed transport properties deviating further from those of stoichiometric NNO.
\begin{figure}
  \includegraphics[angle=0,width=\columnwidth]{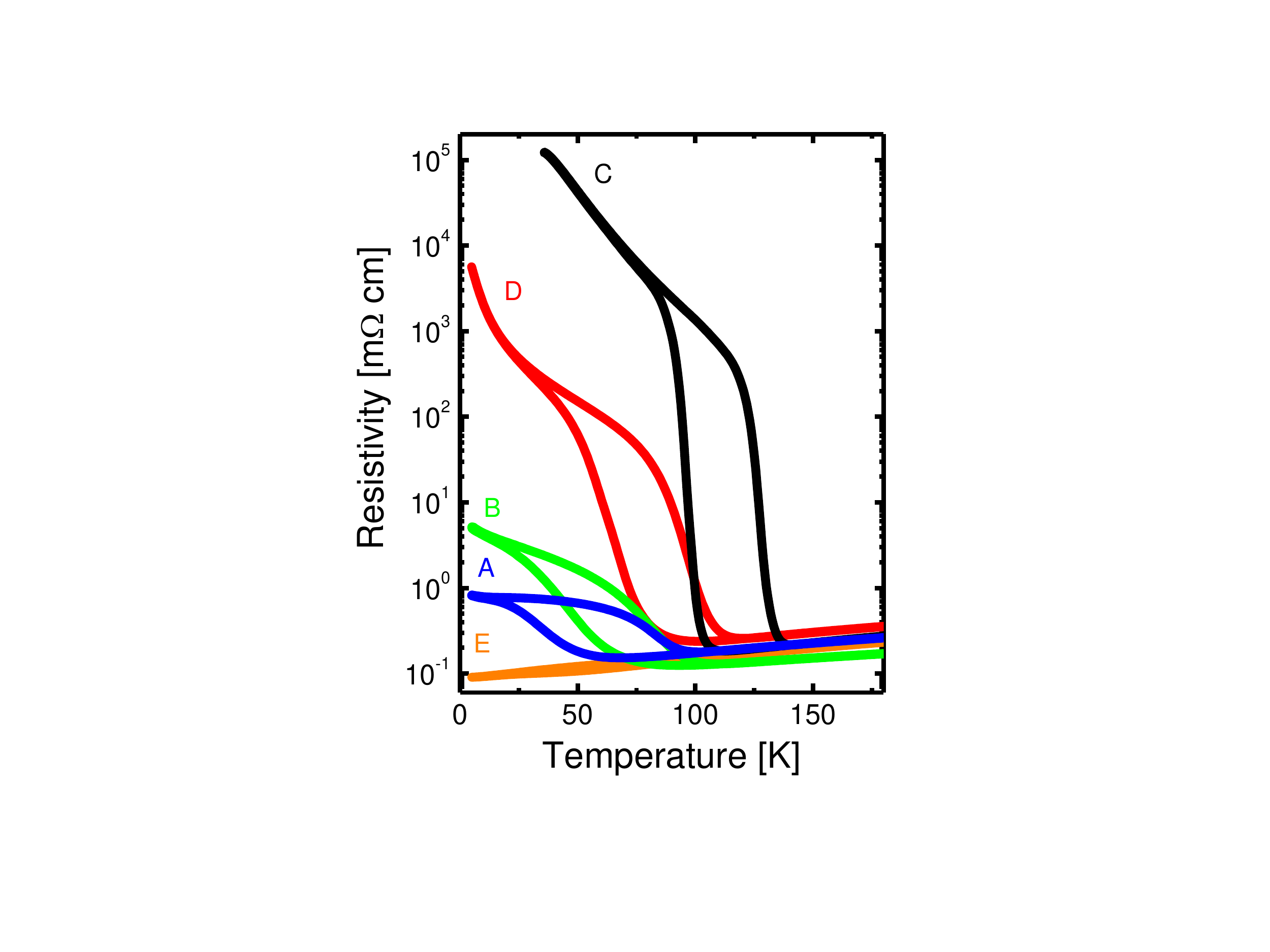}
  \caption{Transport characterization of the five NNO thin films of the series (A-E).}
  \label{fig3}
\end{figure}
To gain insight into film composition and relate it to the transport response, XPS measurements were performed directly after the growth in a UHV chamber connected to the PLD setup, in order to assess the relative cation content for each film. Since XPS is a surface sensitive technique the very small thickness value of the NNO thin films guarantees the complete access to their entire volume (probing depth $ca.$ 5\,nm). Electrostatic charging effects hindered the absolute determination of the core level binding energies associated to the core level peaks of all recorded spectra. As a result, the latter are shown in a relative bending energy (B.E.) scale. As an example, Figure \ref{fig4}a shows the survey photoelectron spectrum obtained for film A where the expected core levels for Nd, Ni and O are labeled. The carbon peak (C1s) at $ca.$ 285 eV is very weak, confirming the good UHV transfer conditions between the PLD and the XPS chambers. Figures \ref{fig4}b,c show better-resolved spectra for the Ni-2p and Nd-3d core levels for samples A and C. 
\begin{figure}
  \includegraphics[angle=0,scale=0.5]{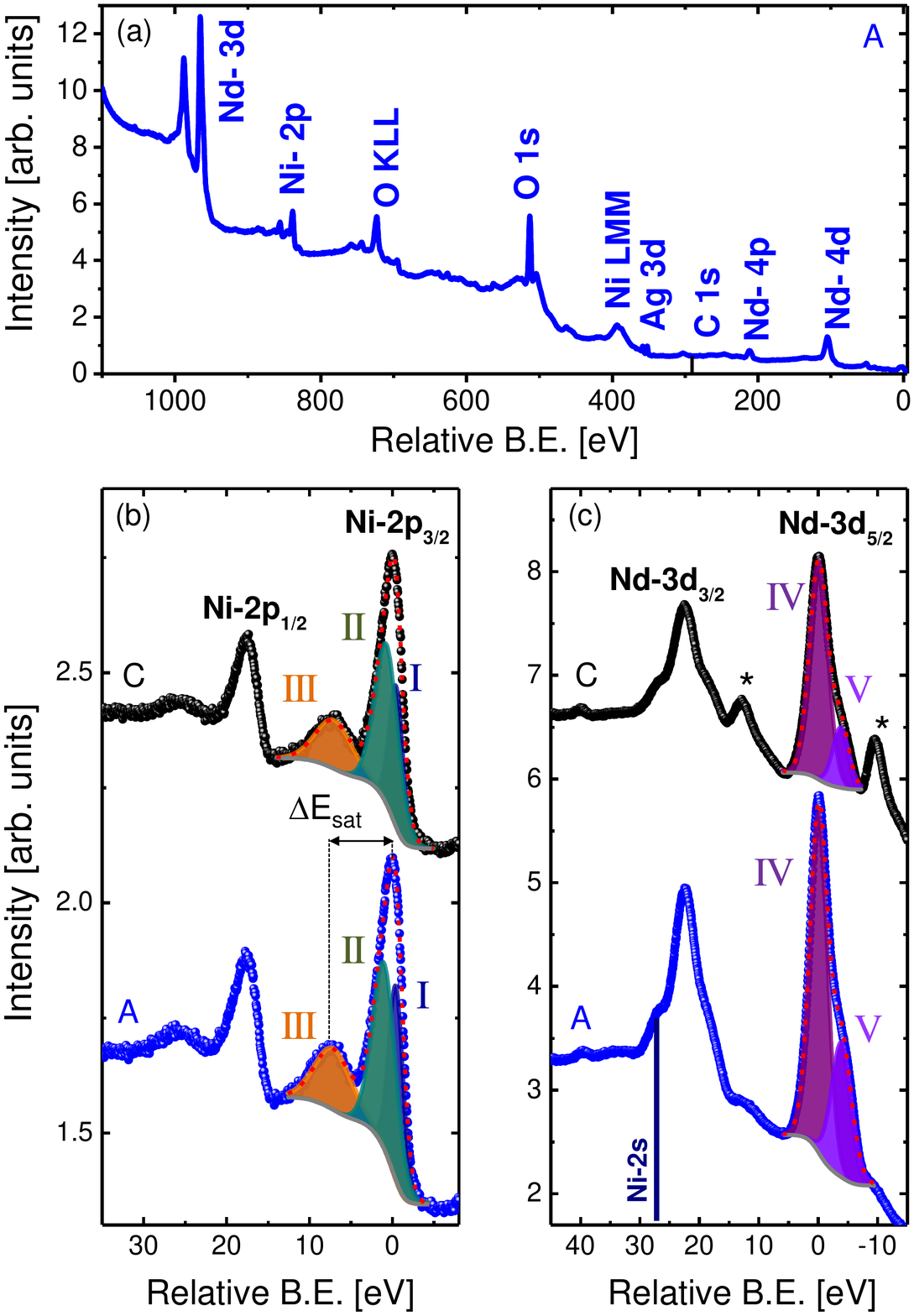}
  \caption{a) Representative XPS survey spectrum (film A). Ni-2p and Nd-3d higher resolved scans for A and C films are depicted in (b) and (c), respectively. Therein the outcome of the peak fitting procedure is also depicted. The * marks a peak due to the Nd deposited onto the sample holder. }
  \label{fig4}
\end{figure}
As expected, the Ni-2p spectra are characterized by the Ni-2p$_{3/2}$ main line, a shake-up satellite separated from the main peak by an energy $\Delta E_{sat}$ and the Ni-2p$_{1/2}$ line with its satellite \cite{Mizokawa_1995, Barman_1994} . For all samples the Ni-2p$_{3/2}$ feature could be properly reproduced by means of two peaks only (labeled as I and II). Thus, we performed a simple fitting (red dotted curve in Figure \ref{fig4}b) of the Ni-2p$_{3/2}$ main line plus satellite and used the total peak area to quantify the Ni content in our films. We analyzed the XPS spectrum of the Nd-3d doublet in a similar way ($cf.$ Figure \ref{fig4}c). Again, we observe two main lines, Nd-3d$_{5/2}$ and the Nd-3d$_{3/2}$ and both exhibit a significant shoulder at lower binding energies. The observed asymmetric shape is characteristic for the Nd-3d doublet in different oxides \cite{UWAMINO198467, atukin2008, Iwanowski2013} . The shake-down satellite next to the main line is due to a charge transfer from the oxygen ligand to the transition metal confirming the charge transfer nature of the NNO \cite{Kohiki_1989} . A quantitative analysis of the Nd-3d$_{3/2}$ line is hindered by a superposition with the Ni-2s level. Therefore, only the Nd-3d$_{5/2}$ level has been fitted by features IV and V which represent the main line and the charge transfer satellite, respectively.\\ To estimate the relative Nd/Ni content in each NNO thin film, we calculated the ratio of the fitted areas of the Nd-3d$_{5/2}$ and Ni-2p$_{3/2}$ spectra after weighing with the relative sensitivity factors of 33 and 13.9 for Nd and Ni, respectively (values taken from the CasaXPS library). The Nd/Ni ratio was found to vary in a relatively large range ($i.e.$ 1.4-3.4). As visible in Figure \ref{fig5}a, the out-of-plane lattice parameter varies systematically with the Nd/Ni ratio, and so do the amplitude of the resistivity variation upon crossing the MIT and the temperature at which the transition occurs, $i.e.$ T$_{MIT}$ ($cf.$ Figures \ref{fig5}b,c). The data suggest that with its very high resistivity change, sample C has the optimal Nd/Ni ratio  ($i.e.$ minimal cationic off-stoichiometry).
\begin{figure}
  \includegraphics[angle=0,width=\columnwidth]{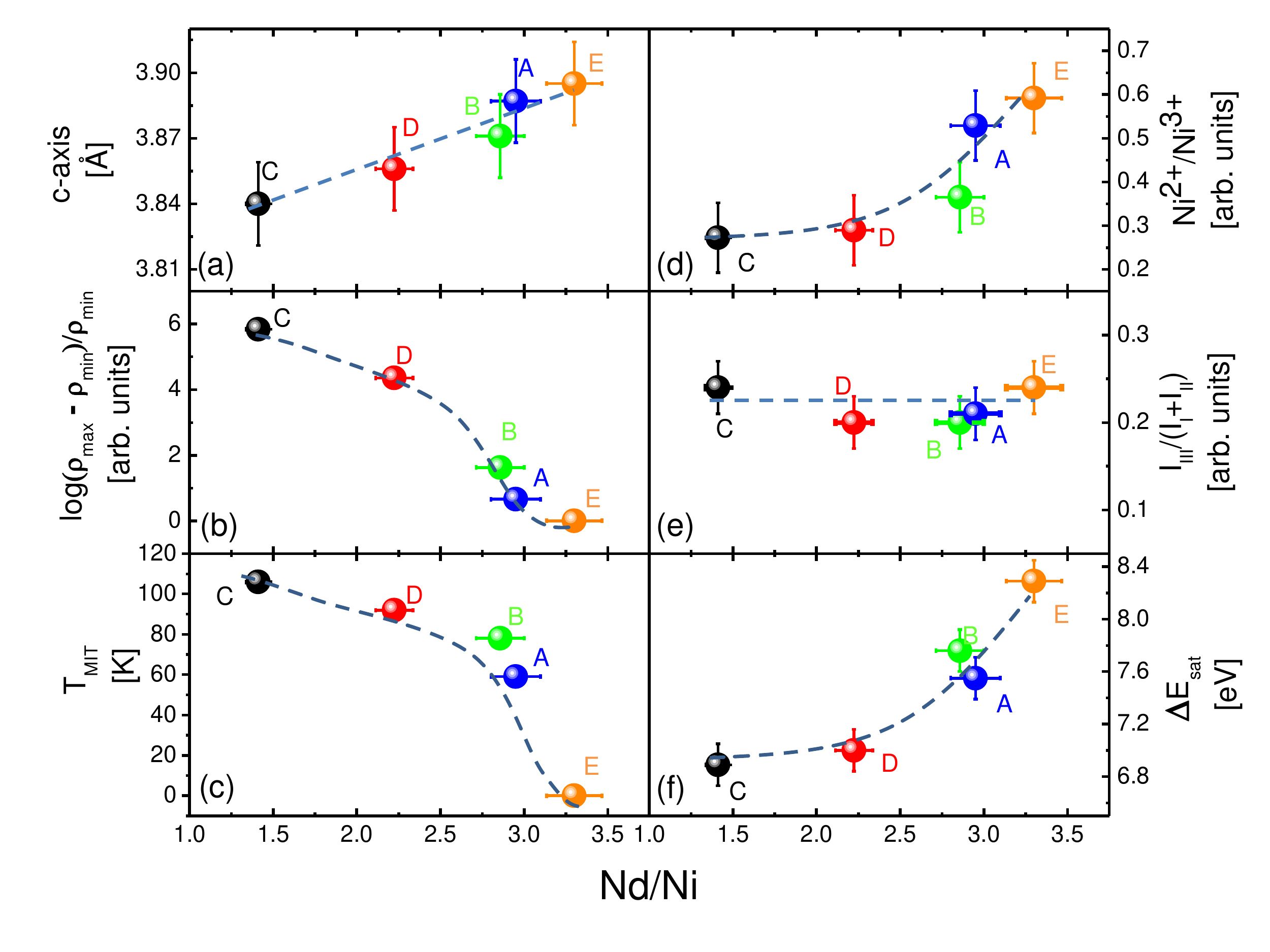}
  \caption{From structural and transport characerization a) out-of-plane lattice paramater, b) resistance change and c) MIT temperature variations as a function of the Nd/Ni ratio. From XPS-based analysis d) Ni$^{2+}$/Ni$^{3+}$, e) covalency and f) $\Delta E_{sat}$ changes as a function of Nd/Ni ratio. For the latter, the error bars of $ca.$ 10\,\% are estimated from a similar analysis relating the Nd-4d and Ni-2p photoelectron spectra.}
  \label{fig5}
\end{figure}
The strong increase of the Nd/Ni ratio for other films implies that they are caracterized by a substantial level of cationic off-stoichiometry, which should strongly affect the valence of the Ni ions. Because we can only estimate the Nd/Ni ratio, at this point we cannot distinguish between Nd excess or Ni deficiency for sample D, B, A and E. Both situations would have opposite effects on the Ni valence: Nd excess would yield a change of the Ni valence from $3+$ towards $2+$, whereas a Ni deficiency would yield a change from $3+$ towards $4+$. As shown by Garcia-Mu\~{n}oz $et\,al.$, both valence changes should result in an enhanced metallicity and a decrease of T$_{MIT}$ \cite{Garc_a_Mu_oz_1995} , as we observe experimentally. \\
To gain insight into the Ni valence, we analyze in more detail the Ni-2p$_{3/2}$ main line ($cf.$ Figure \ref{fig4}b). Following Carley $et\,al.$, one can interpret peaks I and II as related to two different oxidation states ($i.e.$ Ni$^{2+}$ for peak I and Ni$^{3+}$ for peak II)\cite{Carley1999} . For all samples the features I and II are separated in energy by $ca.$ 1.5\,eV and this can explain the absence of Ni$^{4+}$ since, in this case, the energy separation between the 3+ and 4+ oxidation states of Ni is of $ca.$ 3\,eV in an octahedral coordination as reported by Gottschall $et\,al.$ \cite{Gottschall} . Figure \ref{fig5}d shows the relative areas of the Ni$^{2+}$ and Ni$^{3+}$ peaks: at low Nd/Ni ratio, the Ni$^{2+}$ content is the weakest, and it increases for higher Nd/Ni ratios. This implies that the cationic off-stoichiometry in samples D, B, A and E is due to Nd excess. This result is consistent with previous reports of cationic off-stoichiometry in NNO films using Rutherford backscattering spectrometry \cite{Breckenfeld_2014} . We note that even sample C shows some content of Ni$^{2+}$. In view of the excellent transport properties of this sample, this small deviation from pure $3+$ valence probably only reflects slight oxygen deficiency at the film surface rather than a homogenous electron doping due to more severe composition problems.\\
Further information can be extracted from the Ni-2p$_{3/2}$ spectra. In particular, the amplitude of the satellite (feature III) with respect to the main line (I+II) is related to the covalent character of the Ni state. As demonstrated by several studies, in $R$NO compounds Ni cannot be simply described by a 3d$^7$ (purely ionic) state, but by a mixture of 3d$^7$ and 3d$^8\underbar L$ where $\underbar L$ stands for a ligand (oxygen) hole ($cf.$ Figure \ref{fig6}a). The most recent data suggest that the 3d$^8\underbar L$ character strongly dominates over the 3d$^7$ one, highlighting the highly covalent character of these compounds \cite{Bisogni2016} . In addition, the level of covalency ($i.e.$ the relative weight of the 3d$^8\underbar L$ part of the Ni wavefunction, related to the number of holes in the oxygen band), has been shown to decrease upon reducing the rare-earth size, so that LaNiO$_3$, the most metallic compound in the series, is also the most covalent \cite{Freeland201656} . Mizokawa $et\,al.$ have analyzed the Ni-2p XPS spectrum of PrNiO$_3$ in the light of cluster-model calculations \cite{Mizokawa_1995} . Their results indicate that the satellite is a signature of the covalent character. Figure \ref{fig5}e shows the relative amplitude of the satellite peak with respect to that of the main line. If any, the variation over the films series is small, suggesting that the covalency is roughly the same for all samples, $i.e.$ it is conserved irrespective of the Ni$^{2+}$ content. This result is consistent with our observation of a preserved covalency after electron transfer into a nickelate from a rare-earth titanate film \cite{Grisolia2016} .  \\
\begin{figure}
  \includegraphics[angle=0,width=\columnwidth]{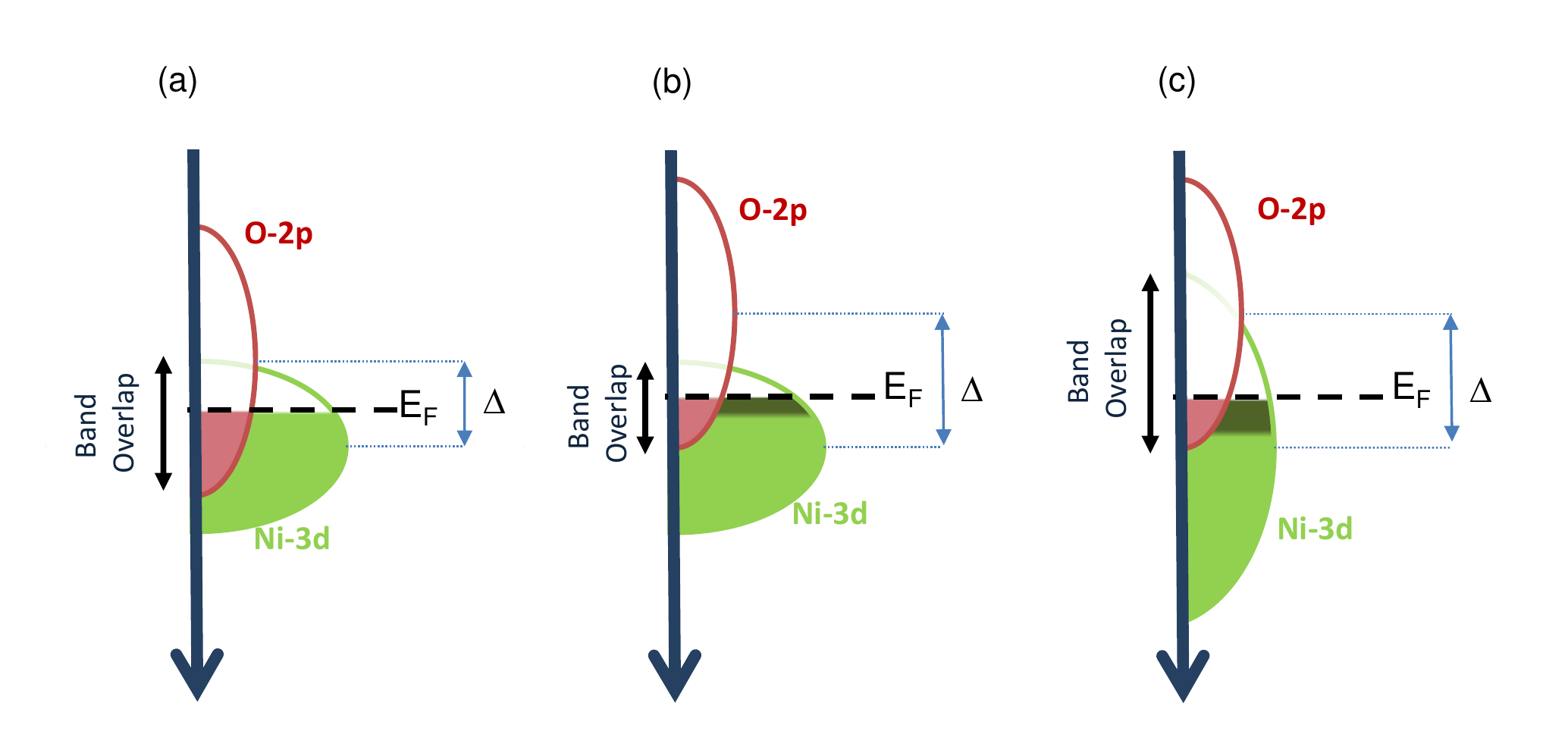}
  \caption{Band picture for the doping effect. In a) the negative charge transfer picture that is modified by adding electrons in b) and allowing the enlargement of the Ni-3d bandwidth in c).}
  \label{fig6}
\end{figure}
Additionally, from the energy separation between the Ni-2p$_{3/2}$ main line and feature III ($i.e.$ $\Delta E_{sat}$ as indicated in Figure 4b), we can obtain further important information related to the transport properties of the NNO series. Indeed, $\Delta E_{sat}$ depends strongly on the Ni 3d-O 2p hopping integral $t_{dp}$, that increases for more metallic nickelates such as LaNiO$_3$ \cite{Barman_1994} . Comparing Figure \ref{fig5}f with Figures \ref{fig5}b and \ref{fig5}c, one notes that $\Delta E_{sat}$ shows a systematic correlation with the suppression of the MIT as the Nd/Ni ratio increases. A relatively high value of 8.3\,eV is obtained for the NNO film E which shows a completely metallic character, close to the value of $ca.$ 9\,eV reported for LaNiO$_3$, which is intrinsically metallic \cite{Barman_1994} .\\
To summarize this analysis, the XPS data suggest that an increase in the Nd/Ni ratio causes an electron doping of the NNO film (change of the Ni valence from 3+ towards 2+), also associated with a suppression of the MIT. For all doping levels, the degree of covalency is globally preserved, but the hopping integral $t_{pd}$ increases with doping. These latter points seem contradictory because maintaining the covalency after electron transfer implies creating more holes in the O-2p band (that is, shifting the O-2p band up in energy with respect to the Ni-3d band), while an increase of $t_{pd}$ implies a stronger overlap between the O-2p and Ni-3d bands. 
This is illustrated by Figure \ref{fig6}b, where the O-2p shifts up in order to preserve the covalency level upon adding electrons. Clearly in this case, the overlap between the O-2p and Ni-3d bands, and thus $t_{pd}$, decrease. However, these two apparently contradictory observations can be reconciled by assuming that as the Nd/Ni ratio increases the Ni-3d bandwidth is enhanced, as illustrated by Figure \ref{fig6}c. This enhancement would reflect a straightening of the Ni-O-Ni bonds at higher Nd/Ni ratio. From our XRD spectra, we inferred that samples with a higher Nd/Ni ratio have a larger out-of-plane parameter ($cf.$ Figure \ref{fig5}a) and volume than the more stoichiometric ones. This is compatible with the increase of the (out-of-plane) Ni-O-Ni bond angles towards 180$^{\circ}$ suggested from the XPS analysis and, consistent with the results of Breckenfeld $et\,al.$ \cite{Breckenfeld_2014} .\\
Beyond the influence of the experimentally observed Nd excess on the electronic properties of the NNO films grown from a mixed-phase target, it proved difficult to provide a satisfactory explanation for the deviation of the films composition from stoichiometry. PLD is a non-equilibrium process, and using mixed phase targets may result in preferential ablation of one component. Indeed, XRD measurements attested the presence of Ni metal peaks onto the surface of the mixed-phase target after the growth of sample E, most probably due to the iterated interaction with the laser (pattern not shown for the sake of simplicity). Even higher fluence values could not improve the reproducibility issue, while deteriorating the structural properties of the obtained NNO thin films. Here, we could not identify a clear trend of the film composition with the target ablation history. To address the issue of preferential ablation in more detail, $\textit{in-situ}$ insights into the plume composition and the ablation dynamics would be required \cite{Orsel2015} , but which lie beyond the scope of this paper.
To resolve this non-stoichiometry issue, we employed a pure NdNiO$_3$ phase target for the growth of a new NNO thin film series as discussed in the next section.

\section{Growth from the single-phase target}
\begin{figure}
  \includegraphics[angle=0,width=\columnwidth]{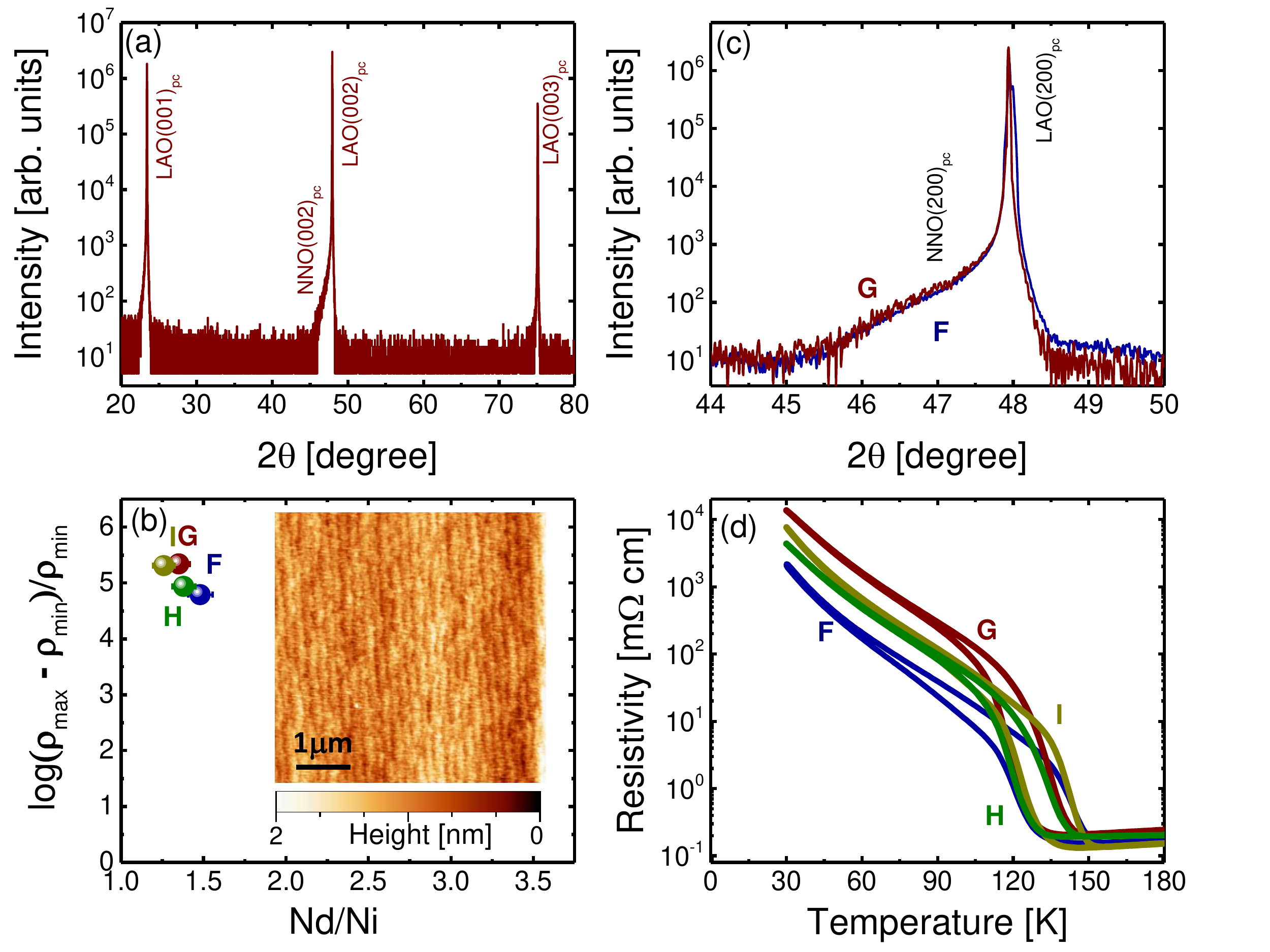}
  \caption{a) Representative XRD pattern of the NNO thin film obtained from a single-phase target with no spurious phases. A high degree of reproducibility is illustrated in b) where the resistivity change is shown as a function of the Nd/Ni peak area ratio obtained from the XPS-based analysis. The inset shows the smooth AFM image. c) Similar out-of-plane lattice parameter for the NNO films exhibiting the largest (G) and the lowest (F) resistance change. d) Transport characterization for the four NNO thin films of the series (F-I). 
  \label{fig7}}
\end{figure}
In this section, we report on the properties of a series of NNO thin films (labeled as F, G, H and I) consecutively grown from a single-phase NdNiO$_3$ target. The XRD pattern of the latter showed no Ni metal peaks at the fluence value of 2\,J/cm$^2$ as before and, all the NNO thin films were grown with the same parameters as for the previous series. All the films were single phase ($cf.$ Figure \ref{fig7}a) exhibiting a smooth and step-terraced morphology ($cf.$ inset of Figure \ref{fig7}b), and with no evidence of the out-of-plane lattice parameter variation ($cf.$ Figure \ref{fig7}c). Importantly, the XPS spectra were very comparable for all samples. In particular the Nd/Ni ratio was in the 1.2\,-\,1.5 range (similarly to sample C from the previous series), $cf.$ Figure \ref{fig7}b. The Ni$^{2+}$/Ni$^{3+}$ peak area ratio was also similar to that of sample C, and $\Delta E_{sat}$ was constant for all samples. Accordingly, all these new samples showed a very similar transport response, with a well-defined MIT and a large resistivity change $cf.$ Figure \ref{fig7}d. This clearly demonstrates a far better degree of reproducibility with the single-phase target (at least up to four consecutive growth experiments).

\section{Conclusion}
Summarizing, we reported on the effect of the ceramic target composition for a physical vapor deposition technique ($i.e.$ PLD), with respect to the reproducibility of the structural, electronic and transport properties of epitaxial NdNiO$_3$ thin films. Initially, the laser fluence value emerged as an important parameter to consider in order to reduce effects of differential ablation process of the target elements and, hence, to reduce the chance of obtaining cationic off-stoichiometry of the desired oxide material. However, even at high fluence an acceptable level of reproducibility could not be achieved when using a mixed-phase target. Indeed, XPS-based analysis revealed a strong off-stoichiometry of the thin films (Nd-excess), causing a reduction of the Ni valence from 3+ towards 2+ and a concomitant enhancement of the Ni-3d bandwidth, translating into a progressive suppression of the metal-insulator transition and a transition to a fully metallic state. \\
 On the contrary, a high degree of reproducibility was demonstrated when a stoichiometric target of a nominal NdNiO$_3$ phase was employed for the growth experiments. In this case the XPS-based analysis indicated a good level of stoichiometry for the NdNiO$_3$ thin films, while transport data revealed well-defined MIT of comparable temperature and amplitude for all samples. \\
Our study indicates that reproducibility issues in PLD growth can be a direct consequence of the ablation process (Laser+Target system). In the future, it would be instructive to perform an appropriate study of the plume dynamic during laser ablation and, check for example, in which way the plume is formed depending on a differential scattering regime of the species present in it, while the morphology/chemistry of the ceramic target is modified.


%
%

%

\begin{acknowledgments}

The authors thank Prof. Jacobo Santamaria for fruitful discussions and for the careful reading of this manuscript.\\This work was financially supported by the ERC Consolidator Grant N$^{\circ}$615759 'MINT' (DP and MB) the Deutsche Forschungsgemeinschaft (HO 53461-1; postdoctoral fellowship to A.S.) and the French National Research Agency (ANR) as part of the "Investissements d'Avenir" program (Labex NanoSaclay, reference: ANR-10-LABX-0035, AXION) (AB).

\end{acknowledgments}

\providecommand{\latin}[1]{#1}
\makeatletter
\providecommand{\doi}
  {\begingroup\let\do\@makeother\dospecials
  \catcode`\{=1 \catcode`\}=2\doi@aux}
\providecommand{\doi@aux}[1]{\endgroup\texttt{#1}}
\makeatother
\providecommand*\mcitethebibliography{\thebibliography}
\csname @ifundefined\endcsname{endmcitethebibliography}
  {\let\endmcitethebibliography\endthebibliography}{}

\end{document}